\documentclass{elsart}
\usepackage{epsfig}
\begin{document}

\def\ket#1{|#1\rangle}
\def\bra#1{\langle#1|}
\def\scal#1#2{\langle#1|#2\rangle}
\def\matr#1#2#3{\langle#1|#2|#3\rangle}
\def\keti#1{|#1)}
\def\brai#1{(#1|}
\def\scali#1#2{(#1|#2)}
\def\matri#1#2#3{(#1|#2|#3)}
\def\bino#1#2{\left(\begin{array}{c}#1\\#2\end{array}\right)}

\begin{frontmatter}
\title{Thermodynamic analogy for quantum phase transitions at zero 
       temperature}
\author[1]{Pavel Cejnar},
\author[2]{Stefan Heinze},
\author[3]{Jan Dobe{\v s}}
\address[1]{Institute of Particle and Nuclear Physics, Charles University,
V Hole{\v s}ovi{\v c}k{\'a}ch 2, 180\,00 Prague, Czech Republic}
\address[2]{Institute of Nuclear Physics, University of Cologne,
Z\"ulpicherstrasse 77, 50937 Cologne, Germany}
\address[3]{Nuclear Physics Institute, Academy of Sciences of the Czech Republic, 
250\,68 {\v R}e{\v z}, Czech Republic}
\begin{abstract}
We propose a relationship between thermodynamic phase transitions
and ground-state quantum phase transitions in systems with variable 
Hamiltonian parameters.
It is based on a link between zeros of the canonical partition 
function at complex temperatures and exceptional points of
a quantum Hamiltonian in the complex-extended parameter space.
This approach is applied in the interacting boson model, where it
is shown to properly distinguish the first- and second-order phase 
transitions.
\end{abstract}
\end{frontmatter}


\section{Introduction}
\label{intro}

Quantum phase transitions (QPT's) are now a well documented phenomenon in 
both lattice \cite{Sachdev} and many-body systems 
\cite{Diep,Feng,Zha,Heiss1,Mor,Rowe,Cas,Cej1,Cej2,Jol1,Jol2,Cej4,Vol,Ema}.
A QPT Hamiltonian usually reads as a superposition of two incompatible
terms,
\begin{equation}
H(\lambda)=H_0+\lambda V=(1-\lambda)H(0)+\lambda H(1)\ ,
\label{Hamg}
\end{equation}
$[H_0,V]\neq 0$,
where $\lambda\in[0,1]$ is a dimensionless control parameter that drives
the system between two limiting modes of motions.
It can be shown that the ground-state (g.s.) average $\langle V\rangle_0
\equiv\matr{\Psi_0(\lambda)}{V}{\Psi_0(\lambda)}$ is a nonincreasing 
function of $\lambda$.
In the QPT situation it evolves in such a way that either $\langle V
\rangle_0$ itself or some of its derivatives change discontinuously (for 
infinite system's size) at a certain critical value $\lambda_{\rm c}$.

Of particular interest are the cases when $\langle V\rangle_0$ drops
to zero and, simultaneously, also $\langle V^2\rangle_0=0$ at the
critical point.
Then the g.s. wave function $\ket{\Psi_0(\lambda)}$ gets fixed
for all $\lambda\geq\lambda_{\rm c}$ and $\langle V\rangle_0$ plays 
a role of an \lq\lq order parameter\rq\rq\ whose value (zero or nonzero) 
distinguishes two quantum \lq\lq phases\rq\rq\ of the model.
This scenario is realized in various many-body models, mostly 
known from nuclear physics.
Limits $H(0)$ and $H(1)$ are usually connected with collective
or single-particle motions corresponding to spherical and deformed 
nuclear shapes 
\cite{Diep,Feng,Zha,Heiss1,Mor,Cas,Cej1,Cej2,Jol1,Jol2,Cej4},
but they can also represent paired and unpaired fermionic phases 
of nuclei \cite{Zha,Heiss1,Rowe,Vol}, or normal and superradiant modes 
of interacting atom-field systems \cite{Ema}.

Questions often arise concerning the depth to which the term 
\lq\lq phase transition\rq\rq\ can be followed in the direction of
standard thermodynamics.
Or is it only a metaphor?
The g.s. QPT's in the sense of Eq.~(\ref{Hamg}) happen at zero 
temperature and thus have no real thermal attributes.
Nevertheless, as discussed in Refs.\cite{Cej2,Cej4,Vol,Zel}, 
counterparts of some thermodynamic terms can often be derived from 
standard quantum-mechanical expressions. 
A unified thermodynamic-like approach to characterize the QPT 
situation is, however, missing.

In this Letter, we present a method capable to establish the 
thermodynamic analogy for quantum phase transitions on a new,
general basis.
The method relates zeros of the canonical partition function $Z(T)$ 
at complex temperatures $T$ \cite{Yang,Gross,Bor} to exceptional, 
or branch points of Hamiltonian (\ref{Hamg}) in complex-extended 
$\lambda$ plane \cite{Heiss1,Zirn,Shan,Steeb,Heiss2,Heiss3,Heiss4}.
This link is not artificial.
While neither zeros, nor exceptional points occur on the real $T$ 
or $\lambda$ axes in the finite case (this would imply divergences of 
thermodynamic observables or real crossings of quantum levels, which 
are generically forbidden), it is known that their distribution in the 
complex plane determines the key features of the system in the physical 
(real) domain.
Zeros and exceptional points thus play crucial roles also in the 
fundamental theory of classical and quantum phase transitions.
Indeed, places where in the thermodynamic limit the complex zeros of 
$Z(T)$ approach infinitely close to the real $T$-axis can be identified 
with points of classical phase transitions \cite{Yang,Gross,Bor}, while
similar convergence of exceptional points to real $\lambda$ induces 
singular evolution of energies and wave functions, as observed in
$T=0$ quantum phase transitions \cite{Heiss1,Heiss3}.
 
\section{QPT analog of specific heat}
\label{analog}

One obvious way to quantitatively exploit the thermodynamic analogy 
relies on connecting the g.s. energy, $E_0(\lambda)$, as a function 
of the interaction parameter $\lambda$, with the equilibrium value of 
a thermodynamic potential, $F_0(T)$, as a function of temperature $T$
(an alternative link to the inverse temperature $\beta$ would not
result in qualitative differences).
From the relation $\langle V\rangle_0=dE_0/d\lambda$ it follows that if the 
$(l-1)$-th derivative of $\langle V\rangle_0$ is discontinuous at 
$\lambda_{\rm c}$, then derivatives of the g.s. energy are discontinuous 
starting from $d^l E_0/d\lambda^l$ so that the QPT is of the $l$-th order 
\cite{Feng}.
The \lq\lq specific heat\rq\rq\ defined through the second derivative 
of $E_0$ in such a transition (in analogy to standard definition
$C=-T\partial^2 F_0/\partial T^2$)
\begin{equation}
C_1=-\lambda\frac{d^2E_0}{d\lambda^2}=-\lambda\frac{d\langle V\rangle_0}
{d\lambda}=2\lambda\sum_{i>0}\frac{|\matr{\Psi_i}{V}{\Psi_0}|^2}{E_i-E_0}
\label{heat1}
\end{equation}
behaves exactly as expected for a thermodynamic phase transition of the 
same order.
Here, $E_i(\lambda)$ and $\ket{\Psi_i(\lambda)}$ are the $i$-th energy and 
eigenvector, respectively.

This relation can be easily verified \cite{Cej4} in the interacting 
boson model (IBM) \cite{Iach}, where both first- and second-order 
phase transitions between spherical and deformed g.s. shapes are 
present in the parameter space 
\cite{Diep,Feng,Mor,Cas,Cej1,Cej2,Jol1,Jol2,Cej4}.
The model describes shapes and collective motions of atomic nuclei in 
terms of an ensemble of $N$ interacting $s$ and $d$ bosons with angular 
momenta 0 and 2, respectively.
We employ a simplified IBM Hamiltonian
\cite{Cas,Cej1,Cej2,Jol1,Jol2,Cej4}
\begin{equation}
H(\lambda)=(1-\lambda)\left[-\frac{q_{\chi}\cdot q_{\chi}}{N}\right]+
\lambda\,n_{\rm d}
\label{ham}
\end{equation}
where $n_{\rm d}=d^{\dagger}\cdot{\tilde d}$ is the $d$-boson number 
operator, and $q_{\chi}=d^{\dagger}{\tilde s}+s^{\dagger}{\tilde d}+
\chi(d^{\dagger}{\tilde d})^{(2)}$ the quadrupole operator.
In the $N\to\infty$ limit, the order parameter $\langle V\rangle_0/N$
(normalization per boson is necessary to deal with effects of varying
$N$) can be expressed in terms of the g.s. deformation parameter 
$\beta_0$ \cite{Diep}:  
\begin{equation}
\lim_{N\to\infty}\frac{\langle V\rangle_0}{N}=
\frac{5\beta_0^2-4\sqrt{\frac{2}{7}}\chi\beta_0^3+
\left(\frac{2}{7}\chi^2+1\right)\beta_0^4}{(1+\beta_0^2)^2}.
\label{order}
\end{equation} 
For $\chi\neq 0$, the value of $\beta_0$ drops from a nonzero value
$\beta_{0{\rm c}}$ to 0 at $\lambda=\lambda_{\rm c}(\chi)=(4+2\chi^2/7)/
(5+2\chi^2/7)$, indicating a first-order deformed-spherical QPT. 
For $\chi=0$, the value $\beta_0\propto\sqrt{\lambda_{\rm c}-\lambda}$
valid in the left vicinity of the critical point $\lambda_{\rm c}(0)$ 
continuously joins with $\beta_0=0$ valid above $\lambda_{\rm c}$; the
corresponding QPT is of the second order. In both cases $\langle V^2\rangle
/N\to 0$ for $\lambda>\lambda_{\rm c}$ and $N\to\infty$.

\begin{figure}
\epsfig{file=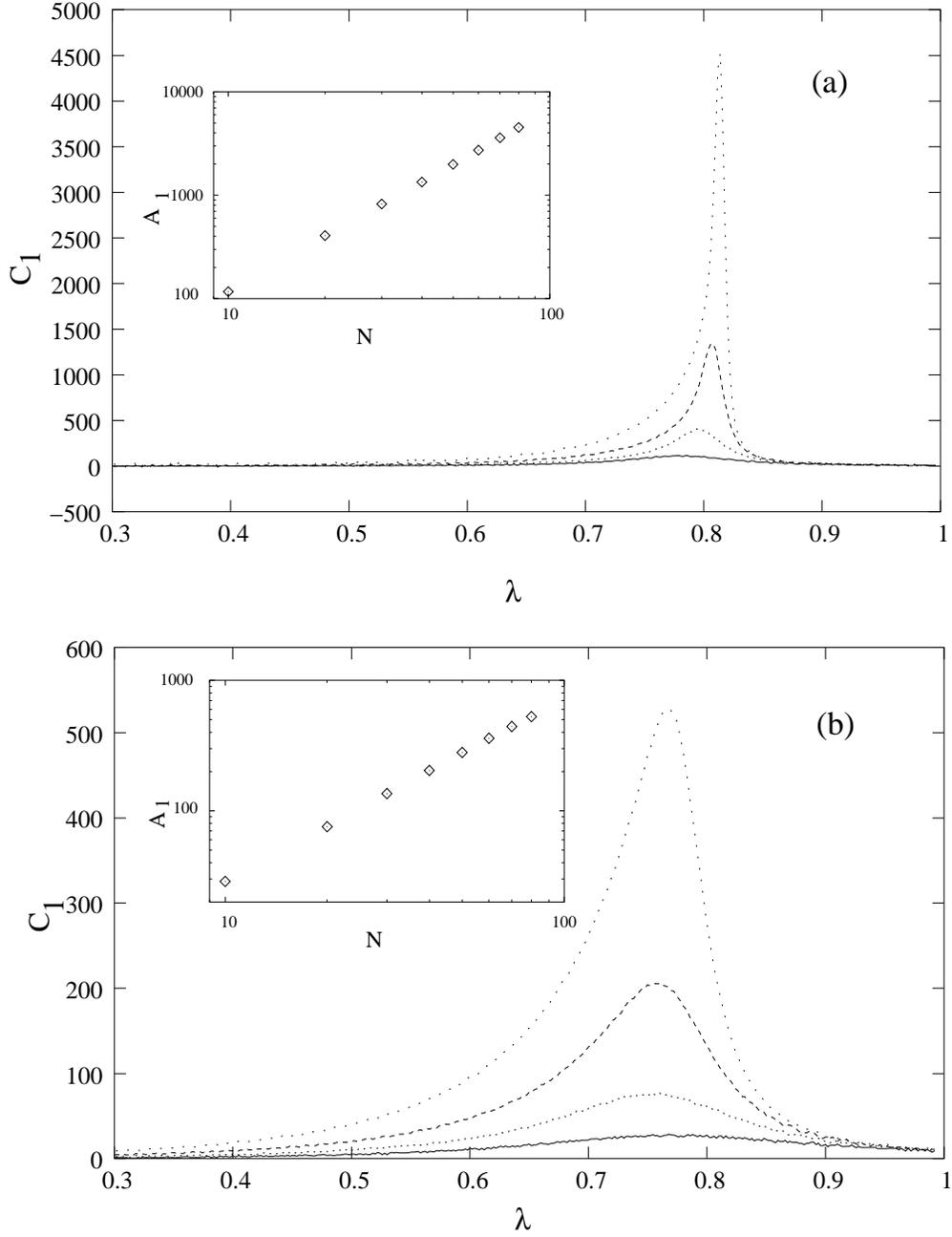,width=\linewidth}
\caption{\protect\small \lq\lq Specific heat\rq\rq\ (\ref{heat1}) in 
the first-order (panel a) and second-order (panel b) QPT of the 
interacting boson model [Hamiltonian (\ref{ham}) with (a) 
$\chi=-\sqrt{7}/2$ and (b) $\chi=0$]. The curves, in order from 
the lowest to the highest, correspond to $N=10$, 20, 40, and 80, 
respectively; the insets show the increase of the maximal value 
with $N$.}
\label{figheat1}
\end{figure}

The dependence of $C_1$ on $\lambda$ in the phase-transitional region 
is shown in Fig.~\ref{figheat1} for various boson numbers $N$. 
Panels (a) and (b) correspond to $\chi=-\sqrt{7}/2$ and 0, respectively.
With an increasing size of the system, the specific heat in panel (a),
normalized per boson, would form a $\delta$-function singularity in 
the $N\to\infty$ limit (first-order QPT), while $C_1/N$ derived 
from panel (b) would develop just a discontinuity at $\lambda_{\rm c}$ 
(second-order QPT).
Note that using the asymptotic $E_0(\lambda)/N\leftrightarrow F_0(T)$ 
correspondence, one can moreover show that the IBM shape-phase diagram 
fully agrees with the classical Landau theory of thermodynamic phase 
transitions \cite{Jol2}.

\section{QPT distribution of exceptional points}
\label{except}

Exceptional points of Hamiltonian (\ref{Hamg}) are points in 
the complex plane of parameter $\lambda$ where various pairs of eigenvalues
of the complex-extended Hamiltonian coalesce.
They are simultaneous solutions of equations $\det[E-H(\lambda)]=0$ and 
$(\partial/\partial E)\det[E-H(\lambda)]=0$, which after elimination
yield the following condition \cite{Zirn,Heiss2}:
\begin{eqnarray}
D(\lambda)=\prod_k D_k(\lambda) & = & 
-\prod_{i<j}[E_j(\lambda)-E_i(\lambda)]^2=0,
\label{disc}\\
D_k(\lambda) & = & \prod_{i(\neq k)}[E_i(\lambda)-E_k(\lambda)].
\label{disck}
\end{eqnarray}
The discriminant $D(\lambda)$ is a polynomial of order $n(n-1)$ in $\lambda$
(where $n$ is the dimension of the Hilbert space) with real coefficients and 
its roots thus occur as $n(n-1)/2$ complex conjugate pairs.
Except at these points, the complex energy $E(\lambda)$ obtained from the 
characteristic polynomial of Hamiltonian (\ref{Hamg}) is a single analytic 
function defined on $n$ Riemann sheets.
The energy labels in Eqs.~(\ref{disc}) and (\ref{disck}) enumerate the
respective Riemann sheet according to the ordering of energies at 
real $\lambda$.
Exceptional points are square-root branch points where the Riemann sheets
are pairwise connected. 
The leading-order behavior on the two connected 
sheets close to the branch point $\lambda_0$ is given by 
$E(\lambda)-E(\lambda_0)\approx a\sqrt{\lambda-\lambda_0}$ (as 
a doubly-valued function), with $a$ being a complex constant 
\cite{Heiss1,Shan,Heiss2}.

The relation of exceptional points to quantum phase transitions has been 
declared several times---see, e.g., Refs.~\cite{Heiss1,Cej1,Heiss3}.
Clearly, an exceptional point located close to the real $\lambda$ axis 
affects the local evolution of the corresponding pair of real energies 
so that the two levels undergo an avoided crossing with accompanying 
rapid changes of wave functions. 
A cumulation of exceptional points close to some real point 
$\lambda_{\rm c}$ thus can give rise to massive structural changes of 
eigenstates, as observed in QPT's.
Although this mechanism was illustrated by several model examples 
\cite{Heiss1,Heiss3}, quantitative determination of the arrangement and 
density of exceptional points needed to trigger a phase-transitional
behavior is still missing.

\begin{figure}
\epsfig{file=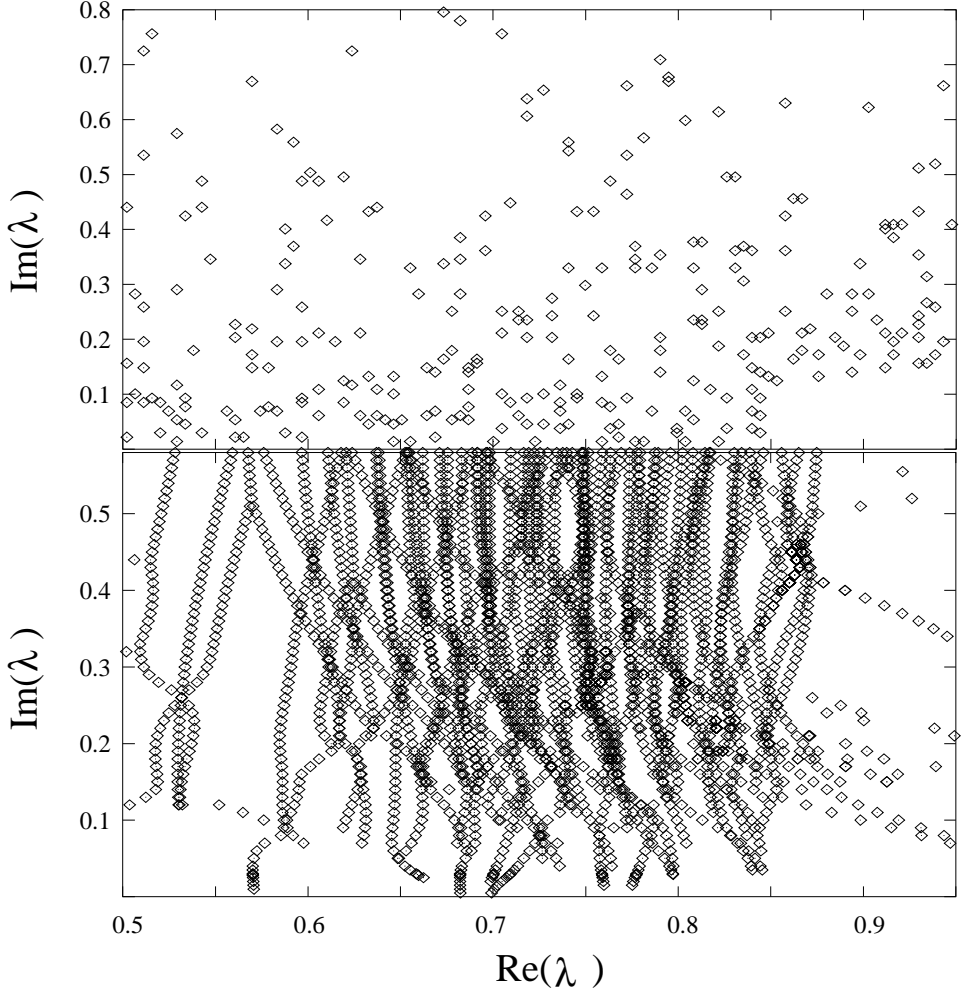,width=\linewidth}
\caption{\protect\small Exceptional points (upper panel) and crossings 
of energy real parts (lower panel) in complex $\lambda$ plane for $J=0$ 
eigenstates of Hamiltonian (\ref{ham}) with $\chi=-\sqrt{7}/2$ and 
$N=20$. Only the lowest 10 levels out of $n=44$ are included in the lower 
panel.}
\label{exc1}
\end{figure}

\begin{figure}
\epsfig{file=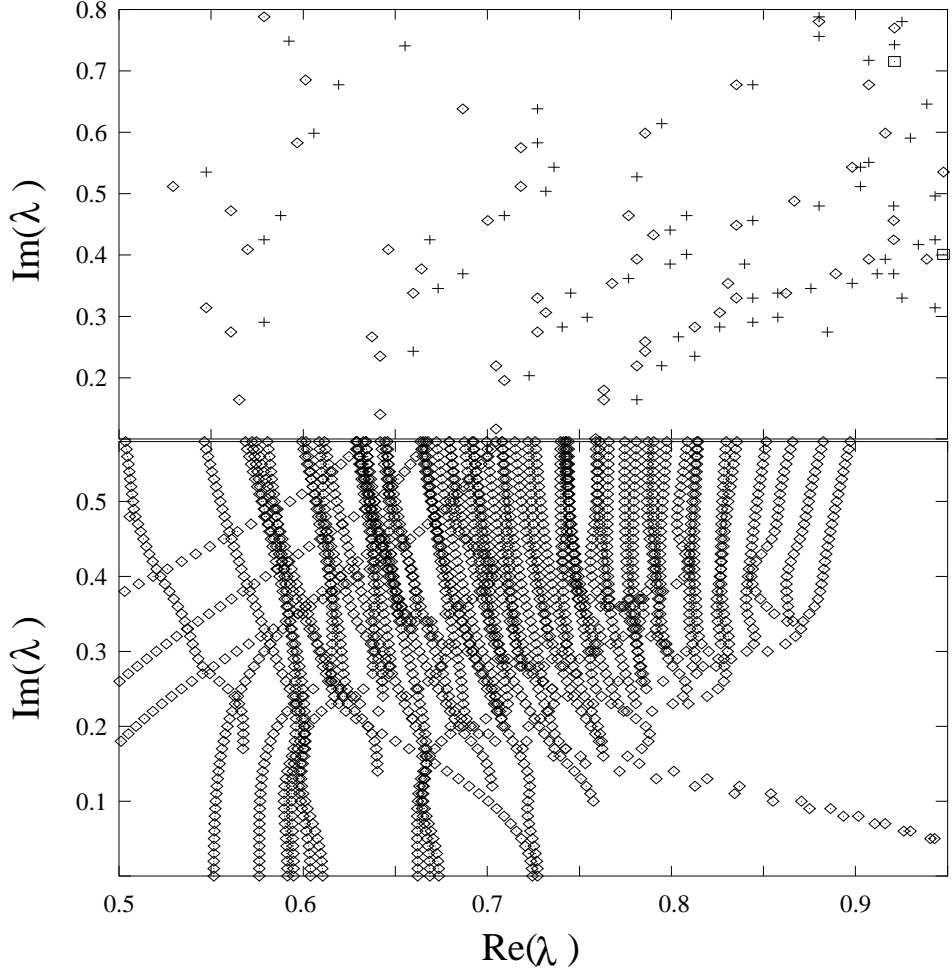,width=\linewidth}
\caption{\protect\small The same as in Fig.~\ref{exc1}, but for $\chi=0$.
In this case, single exceptional points (diamonds) coexist with doubly 
and even quadruply degenerate ones (crosses and rectangles, respectively).}
\label{exc2}
\end{figure}

Exceptional points for levels with angular momentum $J=0$ in the 
vicinity of the IBM first- and second-order phase transitions are 
for $N=20$ shown in the upper panels of Figs.~\ref{exc1} and \ref{exc2}, 
respectively.
The calculation was done using the method described in Ref.\cite{Zirn}.
Lower panels of both figures display---for the lowest ten levels, 
ordered according to ${\rm Re}E_i$---lines where energy real parts 
of two different levels cross, i.e., satisfy ${\rm Re}E_i(\lambda)=
{\rm Re}E_j(\lambda)$.
As follows from the square-root behavior of energies close to crossings, 
the endpoints of these lines coincide with starting points of the 
${\rm Im}E_i(\lambda)={\rm Im}E_j(\lambda)$ lines (not shown here), 
so they represent the exceptional points.

Several exceptional points in Fig.~\ref{exc2} apparently lie on the
real $\lambda$-axis.
This results from the underlying O(5) dynamical symmetry of 
Hamiltonian (\ref{ham}) for $\chi=0$, with the seniority quantum 
number $v$, that survives all the way across the transition and thus 
enables the levels with the same $J$ to cross on the real axis 
\cite{Levi}.
Nongeneric features implied by the additional symmetry at $\chi=0$ 
also include double or even higher degeneracy of some exceptional 
points observed (within the available numerical precision) in 
Fig.~\ref{exc2}.

Broad distributions of exceptional points in Figs.~\ref{exc1} and 
\ref{exc2} do not seem to indicate the QPT behavior at $\lambda
\approx 0.8$, but it must be stressed that these figures contain 
points belonging to various Riemann sheets, while only the 
ground-state Riemann sheet is relevant in the present case.
Unfortunately, to attribute the points to the respective sheets
through the available numerical data becomes problematic already
for relatively small dimensions and the study of effects connected 
with increasing boson numbers seems intractable.

\section{Alternative definition of a QPT specific heat}
\label{analog2}

The above problem can be approached via the relation of the ground-state
related exceptional points to zeros of partition function. 
We will use the correspondence $Z(T)\leftrightarrow D_0(\lambda)$ 
between the partition function and $k=0$ partial discriminant 
(\ref{disck}).
In fact, the square $D_k^2$ of any partial discriminant is a polynomial
with $n-1$ complex conjugate pairs of roots, each of them being 
simultaneously assigned to one other $D_{k'}^2$.
These roots correspond to the branch points located on the $k$-th
Riemann sheet, so that zeros of $D_0$ represent singularities on 
the ground-state Riemann sheet.
A counterpart of the thermodynamic potential $F_0=-T\ln Z$ is 
proportional to the g.s. \lq\lq potential energy\rq\rq,
$U=-\sum_{i>0}\ln|E_i-E_0|$, as obtained in the static Coulomb-gas 
model for quantum energy levels \cite{Dys}.
The divergence of $U$ (zero of $D_0$) implies an infinite 
\lq\lq force\rq\rq\ on the ground state also in the dynamical 
Pechukas-Yukawa model \cite{Pec}. 
For the \lq\lq specific heat\rq\rq\ one obtains (in analogy with 
the standard formula $C=T^2\partial^2\ln Z/\partial T^2+
2T\partial\ln Z/\partial T$)
\begin{equation}
C_2=S\sum_{i>0}\Biggl[\lambda^2\Biggl\{\frac{\frac{d^2E_i}{d\lambda^2}-
\frac{d^2E_0}{d\lambda^2}}{E_i-E_0}-\left(\frac{\frac{dE_i}{d\lambda}-
\frac{dE_0}{d\lambda}}{E_i-E_0}\right)^2\Biggr\}
+2\lambda\left(\frac{\frac{dE_i}{d\lambda}-\frac{dE_0}{d\lambda}}
{E_i-E_0}\right)\Biggr]\ .
\label{heat}
\end{equation}
One can equivalently use $C_2=-S\lambda d^2(\lambda U)/d\lambda^2$, 
which is the same as Eq.~(\ref{heat1}), but with $E_0$ replaced 
by $\lambda U$.
Note that an arbitrary power of $D_0$ used in the thermodynamic 
correspondence would modify just the overall scaling factor $S$.
This factor must also depend on the size of the system (similarly
as the thermodynamic specific heat is normalized to a mass unit) 
and will be discussed later.
Eq.~(\ref{heat}) can be further decomposed \cite{Pec} into a double 
sum of terms containing energy differences (in denominators) and matrix 
elements of $V$ (in numerators), cf. Eq.~(\ref{heat1}).

Although the quantity $C_2$ depends solely on real-$\lambda$ spectral 
observables, its behavior in a QPT is an indirect measure of the 
density of exceptional points on the ground-state Riemann sheet 
beyond (but close to) the real axis.
Consider, for instance, a chain $\lambda_{\pm m}=\lambda_{\rm c}
\pm i{\tilde\lambda_m}$ of $D_0$ zeros along a line 
perpendicular to the real axis.
The $D_0^2$ polynomial is determined (up to a multiplicative 
constant) by the roots and specific heat (\ref{heat}) is for real 
$\lambda$ given by
\begin{equation}
C_2\propto\lambda^2\int_0^{\infty}\frac{\rho({\tilde\lambda})(
{\tilde\lambda}^2-\Delta^2)}{({\tilde\lambda}^2+\Delta^2)^2}
d{\tilde\lambda}
+2\lambda\Delta\int_0^{\infty}{\hskip -2mm}\frac{\rho({\tilde\lambda})}
{{\tilde\lambda}^2+\Delta^2}d{\tilde\lambda}\ ,
\label{cline}
\end{equation}
where $\Delta=\lambda-\lambda_{\rm c}$ and $\rho(\tilde\lambda)$ 
is a density of exceptional points along the $\lambda_{\rm c}+
i{\tilde\lambda}$ line.
This implies that the \lq\lq latent heat\rq\rq\ $Q=\lim_{\epsilon
\to 0}\int_{-\epsilon}^{+\epsilon}C_2(\Delta)d\Delta$ is equal to zero 
if $\rho({\tilde\lambda})$ decreases sufficiently fast when approaching the 
real axis.
In particular, if $\rho({\tilde\lambda})\sim{\tilde\lambda}^{\alpha}$
for ${\tilde\lambda}\to 0$, we obtain the following possibilities:
(i) a first-order QPT, with $C_2\to\infty$ at $\Delta=0$ and $Q$ finite, 
for $\alpha=0$, (ii) second-order QPT, with $C_2\to\infty$ but $Q=0$, for 
$\alpha\in(0,1]$, and (iii) a higher-order QPT, with $C_2$ finite and $Q=0$, 
for $\alpha>1$.
This relation is the same as in standard thermodynamics, where the order 
of a phase transition reflects the density of the $Z(T)$ zeros close to 
a critical temperature $T_{\rm c}$ \cite{Gross}.
It should be stressed, however, that in the QPT case the order deduced 
from $C_2$ cannot be {\it a priori\/} expected to coincide with the order 
determined via $C_1$.
We will focus on this problem in the following, using the IBM.

\begin{figure}
\epsfig{file=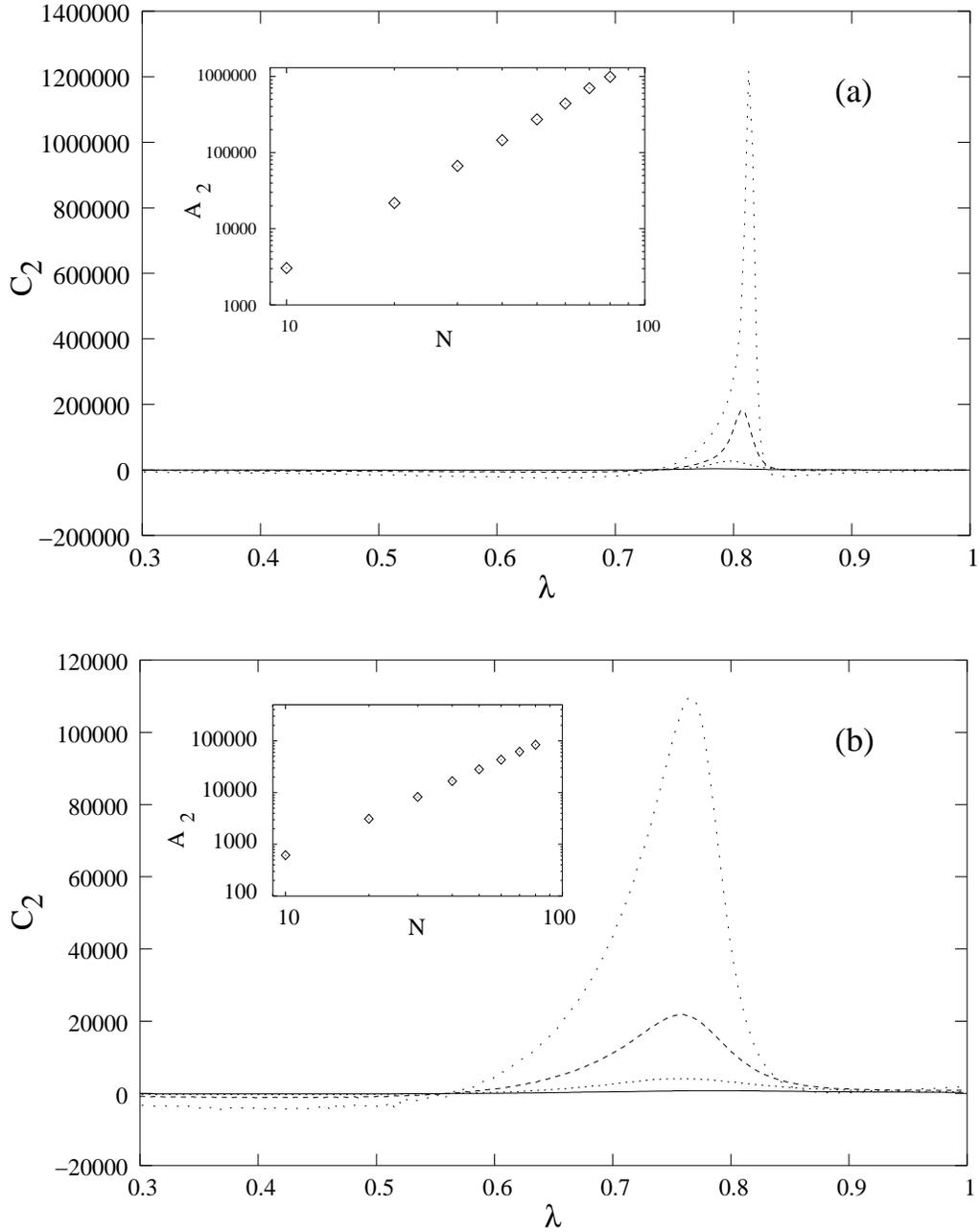,width=\linewidth}
\caption{\protect\small The same as in Fig.~\ref{figheat1} but for 
\lq\lq specific heat\rq\rq\ (\ref{heat}) including all $J=0$ states.} 
\label{figheat}
\end{figure}

The specific heat $C_2$ in the IBM first- and second-order QPT is for $S=1$ 
shown in Fig.~\ref{figheat}, where panels (a) and (b) again correspond to 
$\chi=-\sqrt{7}/2$ and 0, respectively.
All the levels with $J=0$ were included in the sum (\ref{heat}).
We know that for $\chi=0$ some pairs of levels actually cross at real 
$\lambda$ (due to the seniority quantum number $v$).
This implies discontinuities and singularities of the first and second 
derivatives in Eq.~(\ref{heat}), which however cancel out exactly and do 
not affect the $C_2$ shape in Fig.~\ref{figheat}(b).\footnote
{
If $E_i$ crosses with $E_{i+1}$, the discontinuity (singularity) of the 
first (second) derivatives in Eq.~(\ref{heat}) for both levels is the same, 
but with opposite signs, so the sum of both contributions is incremented 
correctly, as if the levels continuously passed the crossing.
}

It is clear that the peaks in panels (a) of both Figs.~\ref{figheat1}
and \ref{figheat} are sharper and higher than those in panels
(b), as indeed expected in the first- and second-order phase
transitions.
The log-log insets in these figures indicate that maximal 
values---$A_1$ and $A_2$---in the $C_1$ and $C_2$ peaks 
exhibit roughly an algebraic increase with the boson number.
The increase is faster for the first-order QPT than for the 
second-order one.
We will show that with a proper normalization to the system's size, 
the values of $A_1$ and $A_2$ corresponding to the second-order QPT 
have a finite $N\to\infty$ asymptotics, while the values of the
first-order QPT, normalized in the same way, diverge.

\begin{figure}
\epsfig{file=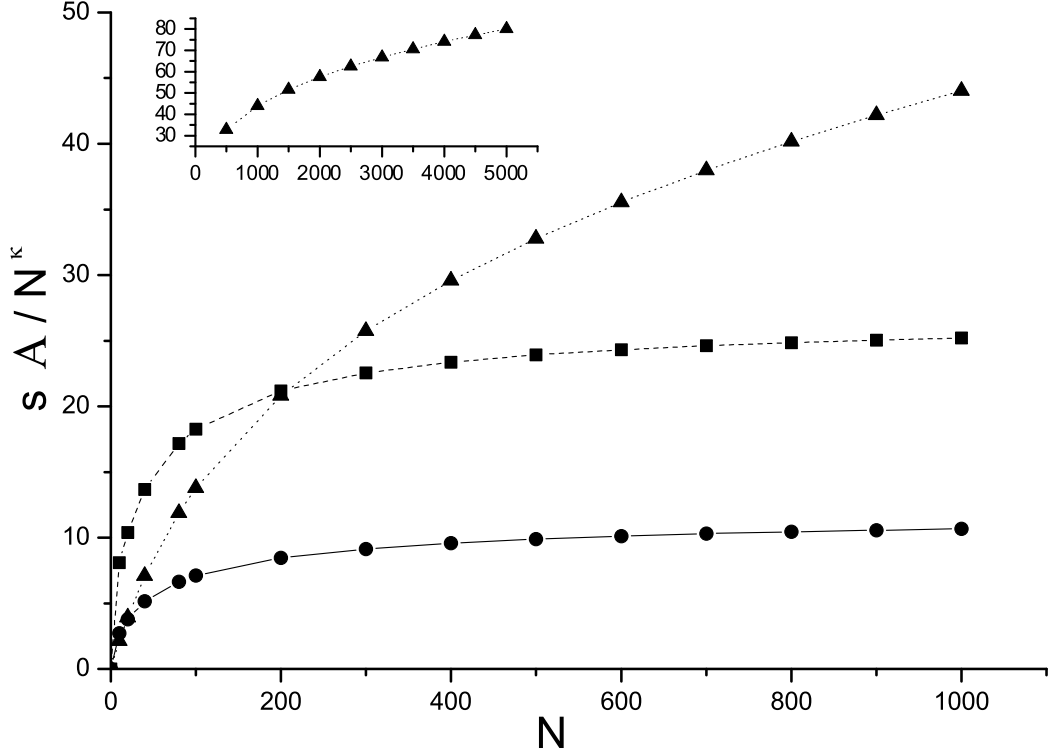,width=\linewidth}
\caption{\protect\small Normalized maximal values $A_1/N$ (dots), 
$s_2 A_2/N^2$ (rectangles), and $s_2' A_2'/N$ (triangles) of 
\lq\lq specific heat\rq\rq\ (\ref{heat1}) and (\ref{heat}) in the IBM 
second-order QPT ($\chi=0$) for very large boson numbers. $A_2$ and 
$A_2'$ were obtained using the whole $J=0$ space or only its $v=0$ 
subspace, respectively. The behavior of $A_2'$ up to $N=5000$ is
shown in the inset. Scaling factors $s_2=10$ and $s_2'=0.5$
were used just to show all data within the same range.}
\label{asymp}
\end{figure}

Fig.~\ref{asymp} shows the dependence of $A_1/N$ and $A_2/N^2$ 
for the second-order QPT on the boson number up to $N=1000$.
The calculation for such high dimensions was enabled by the
underlying O(5) symmetry for $\chi=0$.
Clearly, the curves in Fig.~\ref{asymp} have finite asymptotics.
The behavior of $A_1$ is consistent with the analytic result 
$A_1/N\to12.5$ that can be derived from the $N\to\infty$
limit of the g.s. energy per boson.
The faster increase of $A_1$ in the first-order transition, see 
Fig.~\ref{figheat1}(a), indicates a divergence of $C_1/N$ at 
$\lambda_{\rm c}(\chi)$ for $\chi\neq 0$.

Specific heat $C_2$ in Fig.~\ref{figheat} behaves in a similar
way, but---as follows from Fig.~\ref{asymp}---in this case the 
correct normalization is by $1/N^2$.
This factor reflects the dimension of the subspace of states 
with $J=0$ which grows roughly as $n\sim N^2/12$ for very 
large boson numbers (while the total number of IBM states is 
$\sim N^5/120$).
Since there are $n-1$ pairs of exceptional points on each Riemann 
sheet, the choice of $S=1/(n-1)\propto 1/N^2$ in Eq.~(\ref{heat}) 
normalizes the density $\rho({\tilde\lambda})$ [cf. Eq.~(\ref{cline})] 
in the finite-$n$ case
to a unit integral and plays a similar role as the $1/N$ scaling
of $C_1$ in the previous case.
With this normalization, the specific heat $C_2$ correctly 
distinguishes the IBM first- and second-order phase transitions.

This conclusion is also supported by the analysis of the IBM
classical limit.
Indeed, due to the degeneracy of $\beta_0=0$ and $\beta_0\neq 0$ 
minima of the $N\to\infty$ g.s. energy per boson at $\lambda_{\rm c}$ 
\cite{Diep}, there is an exceptional point on the ground-state 
Riemann sheet that is for the infinite boson number located on 
the real axis, at the first-order QPT.
Consequently, energy denominators in Eq.~(\ref{heat}) diverge
at this point in a way that appears to be faster than $\propto N^2$. 
In contrast, there is no actual ground-state involving degeneracy 
in the asymptotic limit of the second-order transition, although, 
as can be shown, exceptional points come infinitely close to the 
real axis at the critical point.
This results in a slower increase, which is compensated by the 
$1/N^2$ scaling.

For $\chi=0$, we can additionally check the validity of the $S=1/(n-1)$
normalization of specific heat (\ref{heat}) by considering only the 
$v=0$ subspace of $J=0$ states, with the dimension $n\sim N/2$.
Since this subspace contains the ground state and does not mix with
$v\neq 0$ subspaces, one can restrict the sum in Eq.~(\ref{heat})
only to these states and obtain a modified specific heat $C_2'$.
It is expected to have a similar behavior as $C_2$, but with
$S\propto 1/N$. 
Unfortunately, the convergence of the maximal value $A_2'$ to some 
asymptotics is very slow.
The available data for boson numbers up to 5000 (see the inset in
Fig.~\ref{asymp}) can only yield an upper constraint for the power 
in $S\propto 1/N^{\kappa}$, namely $\kappa<1.35$, which is consistent 
with the expectation but not fully conclusive.
Work in this direction is in progress.

\section{Conclusions}
\label{conclus}

We have developed a method capable to indirectly measure the 
distribution of exceptional points of Hamiltonian (\ref{Hamg})
on the ground-state Riemann sheet near the real-$\lambda$ axis.
This distribution plays major role in structural changes of the 
g.s. wave function. 
It is crucial in the determination of zero-temperature quantum 
phase transitions, but numerically difficult to be calculated 
directly for large dimensions.
The method is based on the analogy between generic QPT arrangements 
of exceptional points and similar behaviors of complex zeros of the 
canonical partition function in thermodynamic phase transitions.
The resulting formula (\ref{heat}) for the \lq\lq specific heat\rq\rq\ 
$C_2$ normalized with respect to the relevant dimension, $S\sim 1/n$, 
turned out to be the appropriate measure for the cumulation of
exceptional points near the QPT critical point.
Our approach was tested in the interacting boson model, where 
$C_2$ in the first- and second-order QPT's was shown to behave in 
the same way as standard specific heat in typical thermodynamic 
phase transitions of the respective orders.
We expect that the method is applicable also in other QPT systems, 
even beyond nuclear physics. 
It discloses surprising analogy between standard thermodynamics and 
quantum mechanics of parameter-dependent systems.

\section*{Acknowledgments}
P.C. and S.H. thank Jan Jolie for relevant discussions.
This work was supported by GA{\v C}R and AS{\v C}R under
Project Nos. 202/02/0939 and K1048102, respectively, and
by the DFG under Grant No. 436 TSE 17/6/03.

\end{document}